\documentclass[12pt,preprint]{aastex}
\slugcomment{To appear in the Astrophysical Journal}

\lefthead{Skinner et al.}
\righthead{RY  Tau}

\begin{document}

\title{{\em Chandra} Evidence for Extended X-ray Structure in RY Tau}

\author{Stephen L. Skinner\footnote{Center for Astrophysics and 
Space Astronomy (CASA), Univ. of Colorado,
Boulder, CO, USA 80309-0389; stephen.skinner@colorado.edu},
Marc Audard\footnote{ISDC Data Center for Astrophysics, 
University of Geneva, Ch. 
d'Ecogia 16, CH-1290 Versoix, Switzerland and
Geneva Observatory, University of Geneva, Ch. des Maillettes 51,
CH-1290 Versoix, Switzerland; marc.audard@unige.ch}, and
Manuel  G\"{u}del\footnote{Dept. of Astronomy, Univ. of Vienna, 
T\"{u}rkenschanzstr. 17,  A-1180 Vienna, Austria; 
manuel.guedel@univie.ac.at}}

% The abstract environment prints out the receipt and acceptance dates
% if they are relevant for the journal style.  For the aasms style, they
% will print out as horizontal rules for the editorial staff to type
% on, so long as the author does not include \received and \accepted
% commands.  This should not be done, since \received and \accepted dates
% are not known to the author.
%
% Define symbol \ltsimeq
\newcommand{\ltsimeq}{\raisebox{-0.6ex}{$\,\stackrel{\raisebox{-.2ex}%
{$\textstyle<$}}{\sim}\,$}}
%
% Define symbol \gtsimeq
\newcommand{\gtsimeq}{\raisebox{-0.6ex}{$\,\stackrel{\raisebox{-.2ex}%
{$\textstyle>$}}{\sim}\,$}}

\begin{abstract}
\small{
We report results of a sensitive {\em Chandra} ACIS-S observation 
of the classical  T Tauri star RY Tau. Previous studies 
have shown that it drives a spectacular bipolar
jet whose blueshifted component is traced optically 
along P.A. $\approx$ 295$^{\circ}$ at separations of 
1.$''$5 - 31$''$ from the star. 
Complex X-ray emission is revealed, including a very soft
non-variable spectral component (some of which may originate 
in shocks), a superhot flaring  component 
(T $\gtsimeq$ 100 MK), and faint extended structure near
the star. The structure is visible in deconvolved images
and extends northwestward out to a separation of 1.$''$7,
overlapping the inner part of the optical jet. Image analysis
suggests that most of the extension is real, but some 
contamination by  PSF-induced structure within the central
arcsecond may be present.  The predicted
temperature for a shock-heated jet based on  jet speed
and shock speed estimates from optical measurements is too
low to explain the extended X-ray structure. Either higher
speed material within the jet has escaped optical detection
or other mechanisms  besides shock-heating are involved. 
Alternative mechanisms that could produce higher temperature plasma 
at small offsets to the northwest of RY Tau include magnetic
heating in the jet, hot plasmoids ejected at high speeds, or
X-ray emission from a putative close companion whose presence
has been inferred from {\em Hipparcos} variations.
}
\end{abstract}

% The different journals have different requirements for keywords.  The
% keywords.apj file, found on aas.org in the pubs/aastex-misc directory, 
% contains a list of keywords used with the ApJ and Letters.  These are 
% usually assigned by the editor, but authors may include them in their 
% manuscripts if they wish. 

\keywords{stars: individual (RY Tau) --- ISM: Herbig-Haro objects ---  
stars: pre-main sequence --- stars: winds, outflows --- X-rays: stars}

% That's it for the front matter.  On to the main body of the paper.
% We'll only put in tutorial remarks at the beginning of each section
% so you can see entire sections together.

%\newpage

\section{Introduction}

Jets are  known to be associated with a diverse range of
astronomical objects including active galactic nuclei,
compact binaries, planetary nebulae, brown dwarfs, and
young stellar objects (YSOs).
The mechanisms  responsible for jet launching, collimation, 
and heating are still not well-understood so astrophysical 
jets continue to be the focus of numerous observational, 
laboratory, and theoretical studies.

Jets are thought to be intimately linked to the process of 
accretion. In star-forming regions, collimated jets have
been detected from accreting YSOs
such as embedded protostars, classical T Tauri stars (TTS),
and  Herbig Ae/Be stars.  Jets from YSOs are capable of influencing
the star-formation process by injecting energy, momentum, and associated
turbulence into the surrounding molecular cloud. Spectroscopic studies
suggest that some YSO jets are rotating  (Chrysostomou et al. 2008),
implying that jets  play a role in transporting angular momentum
outward in young star-disk systems. Precessing  jets also 
provide an indirect means of detecting dynamical interactions in 
close protostellar binaries (Su et al. 2007).

YSO  jets have traditionally been identified and studied using 
high-resolution optical, near-IR,  or radio telescopes. But, the improvement
in angular resolution afforded by current-generation X-ray telescopes
has revealed that YSO jets are also capable of emitting soft X-rays
originating in  plasma at temperatures of a few MK. 
X-ray observations are important because they provide information
on physical conditions in the hottest jet plasma that is
not accessible at optical or infrared wavelengths. Only a few
examples of X-ray emitting YSO jets  are known so far. The
most striking example is the classical TTS DG Tau, which shows soft
X-ray emission extending outward along the optical jet axis
to a distance of $\approx$5$''$ from the star 
(G\"{u}del et al. 2005, 2008).  
We report here the first evidence of asymmetric X-ray structure
in the  accreting TTS RY Tau that is visible as faint extension
toward the northwest. The extended structure may be associated
with the inner region of its optically-traced jet, but 
alternative interpretations are also discussed.

\section{RY Tau}

RY Tau is a  classical TTS 
lying in the Taurus dark cloud (d = 134 pc;
Bertout, Robichon, \& Arenou 1999). It is 
highly variable in the optical and its spectral type
is  somewhat uncertain (Holtzman, Herbst, \& Booth 1986).
Its mass is at the high end of the range for TTS.
Recent work suggests a spectral type of F8 III - G1-2 IV,
mass M$_{*}$ = 1.7 - 2.0 M$_{\odot}$, luminosity L$_{*}$ = 12.8 L$_{\odot}$,
and A$_{\rm V}$ = 2.2 $\pm$ 0.2 mag (Schegerer et al. 
2008 and references therein).  The star rotates  rapidly
with $v$sin$i$  = 52 $\pm$ 2 km s$^{-1}$ (Petrov et al. 1999).
Previous optical and IR studies have shown that RY Tau  
has an accretion disk and is accreting at  a  rate 
$\dot{\rm M}_{acc}$ = 6 ($\pm$3) $\times$ 10$^{-8}$
M$_{\odot}$ yr$^{-1}$ (Schegerer et al. 2008; Angra-Amboage
et al. 2009, hereafter AA09). The accretion is accompanied by mass-loss in the 
form of a spectacular bipolar jet known as HH 938
(St.-Onge \& Bastien 2008, hereafter SB08) and a wind
(G\'{o}mez de Castro \& Verdugo 2007).
H$\alpha$ images show several  jet knots extending 
outward to a separation of 31$''$ from the star along
P.A. $\approx$ 295$^{\circ}$ (measured east from north), and
traced inward to a separation of $\approx$1.$''$5 (SB08).
YSO jets  traced so close to the star are rare.
The fainter counterjet is optically visible in the opposite 
direction out to 3.$'$5 from the star. 

RY Tau is suspected to be a binary 
on the basis of nonuniform photocenter motions in
{\em Hipparcos} observations.
Analysis of the {\em Hipparcos} variations initially gave a
direction P.A. = 316.6$^{\circ}$ $\pm$ 37.6$^{\circ}$ 
toward the putative companion, but a later reanalysis 
found P.A. = 304$^{\circ}$ $\pm$ 34$^{\circ}$
and a companion separation   $\geq$23.6 mas
(Bertout et al. 1999). But, no companion
has yet been found (Leinert et al. 1993; Schegerer et al. 2008; 
Pott et al. 2010). The P.A. of the putative companion is close to that 
of the optical jet and it has been suggested that the
jet  may have been responsible for the {\em Hipparcos}
variations (AA09).

A previous {\em Chandra}
ACIS-S/HETG  observation of the TTS HDE 283572 (ObsId 3756)
captured RY Tau as a bright off-axis X-ray source 
(Audard et al. 2005), but the archived data cannot
be used to search for jet emission close to the star
because of off-axis image distortion. This motivated us to
obtain a higher-quality {\em Chandra} image  with 
RY Tau placed on-axis.

\section{Chandra Observation}

%%For on-axis point sources, {\em Chandra} ACIS-S
%%is capable of focusing the emission from the central star
%%into a core of radius $\approx$0.7$''$ (80\% encircled power;
%%POG Fig. 6.7). [NOTE: this disagrees with statement in
%%next section.] Thus, any extended jet emission that might
%%be present can be traced down to a separation of $\approx$1$''$.

The {\em Chandra} observation (ObsId 10991) occurred  on 
2009 December 31 from 05:31 - 22:15 TT. 
The exposure live time was 55,766 s. 
Exposures were obtained using the ACIS-S (Advanced CCD 
Imaging Spectrometer) array  in faint  timed-event mode.
ACIS-S was restricted to the standard 1/4 subarray  with only
one chip (S3) enabled.  This reduces the field-of-view to
2.1$'$ $\times$ 8.4$'$ but allows 
a shorter 0.8 s frame readout time (compared to 3.2 s for 
the full ACIS-S array), thus mitigating  photon 
pileup\footnote{Pileup occurs when two or more photons are detected
as a single event. Severe pileup can  broaden the 
source image and  artificially harden the X-ray spectrum.
For more details, see::
http://cxc.harvard.edu/ciao/why/pileup\_intro.html ~.}
for this bright source (Sec. 4.2).  RY Tau was placed 
at the nominal ACIS-S 1/4 subarray aimpoint. 
%% ACIS-S has a 90\% encircled energy
%% radius of R$_{90}$ $\approx$ 1.$''$96 at 2 keV
%%for an on-axis point source
%%\footnote{http://cxc.harvard.edu/cal/Acis/Cal\_prods/psf/eer\_on.html}. 

The data were  analyzed using standard science
threads in CIAO\footnote{Further information on 
{\em Chandra} Interactive
Analysis of Observations (CIAO) software can be found at
http://asc.harvard.edu/ciao .}  version 4.1.2
and calibration data from CALDB version 4.1.4.
The default ACIS pixel randomization applied during standard
processing slightly broadens the point-spread-function (PSF)
and was removed for image analysis. However, we have 
compared our imaging results
(Sec. 4.3) with the default pixel randomization retained and 
there are no significant differences.
CIAO {\em specextract} was used to extract  source and background 
spectra (background is negligible).
%%response matrix files (RMFs) and auxiliary response files (ARFs).
%%We used a circular region of radius 6 pixels (r = 2.$''$95)
%%to extract source events for spectral and light curve analysis.
%%Background was extracted from adjacent source-free regions. 
%%Background is negligible, contributing only 0.11\% 
%%of the total  counts in the 0.3 - 8 keV range
%%inside a circular region of 
%%radius r = 6 pixels (r =  2.$''$95) centered on the source.
The tool $acisreadcorr$ was used to remove a readout streak
running east-west through the source along 
P.A. = 91$^{\circ}$/271$^{\circ}$. 
Spectral fitting was undertaken with XSPEC vers. 12.4.0
\footnote{http://heasarc.gsfc.nasa.gov/docs/xanadu/xanadu.html.}.

%%\vspace*{0.7cm}
\section{Results}

\subsection{The X-ray Field Near RY Tau }

Figure 1 shows the {\em Chandra} broad-band image in the 
vicinity of RY Tau. The bright emission from RY Tau dominates
the field, but 5 fainter sources within $\approx$1$'$
of it were found by the CIAO wavelet detection tool
{\em wavdetect}. Their properties are summarized in Table 1.
No optical or near-IR counterparts were found for  any of
these field sources in the {\em HST} Guide Star Catalog (GSC)  
or 2MASS catalog. 

A direct comparison of the X-ray source positions
(R.A., Dec.)  in Table 1 with the optical knot positions in
SB08 cannot be made because the tabulated SB08 optical positions 
have large systematic errors. Specifically, we note that
the reference position of RY Tau given in Tables 1 and 2
of SB08 is offset 7.$''$0 eastward and 
0.$''$5 northward from its {\em HST} GSC v. 2.3.2 position 
(= {\em HST} J042157.41$+$282635.48). However, a comparison
of the optical knot offsets relative to the star given by
SB08 with the {\em Chandra} source offsets (Table 1)
reveals a possible match between {\em Chandra} source no. 2
and optical knot Ha6.    The  offset of X-ray 
source no. 2 (CXO J042155.93$+$282647.55) from the RY Tau X-ray
centroid is 23.$''$1, and its offset relative to the {\em HST}
GSC position of RY Tau is 22.$''$9. These values are close
to the 22.$''$2 offset of knot Ha6 from the reference position
of RY Tau listed in Table 2 of SB08, but this comparison does
not take into account any proper motion that might have 
occurred in knot Ha6 since the SB08 observation on 
25 February 2005.  The offset of  X-ray
source no. 2 from the RY Tau X-ray peak is along P.A. = 302$^{\circ}$ 
which is comparable to the value P.A. = 293$^{\circ}$ computed for the 
Ha6 optical knot direction (Table 2 of SB08). Also of
potential interest is the fainter X-ray source 
no. 4 (CXO J042200.09$+$282613.39),  
which lies exactly in the opposite direction along P.A. = 122$^{\circ}$.
Since X-ray sources no. 2 and no. 4 have relatively high median
photon energies (Table 1), deep follow-up optical observations
will be needed  to determine whether they are chance superpositions
of foreground or background objects near the jet axis, or
shocked features in the jet.

\subsection{X-ray Variability and Pileup Estimates} 
The 0.2 - 8 keV broad-band light curve of RY Tau
(Fig. 2-top) is clearly variable, 
showing an increase in count rate by a factor of
$\sim$2 during the first 10 ks, followed by a slow 
decay with an e-folding timescale $\tau \approx$ 10 ks.
The variability is  associated with the harder emission 
above $\approx$1 keV and is likely due to  
magnetic activity. If we consider only the softest
emission in the 0.2 - 1.0 keV range (910 events)
where shock emission could contribute significantly,
little or no variability is found (Fig. 2-bottom).
Specifically, the Bayesian-method CIAO tool {\em glvary}
(Gregory \& Loredo 1992, 1996) gives a variability 
probability P$_{var}$ = 0.05 (0.2 - 1.0 keV). But, if 
the energy range is increased to 0.2 - 1.5 keV, 
{\em glvary} gives  P$_{var}$ = 0.999, so the emission
above 1 keV is definitely variable.

Even though a flare occurred, we were fortunate to 
avoid any severe pileup as a result of the judicious
use of a 1/4 ACIS-S subarray and its shorter 0.8 s
frame readout time. We estimated the pileup fraction
using the CIAO {\em pileup\_map} tool and the 
{\em PIMMS}\footnote{http://asc.harvard.edu/toolkit/pimms.jsp}
simulator. Moderate pileup of $\approx$10\% - 13\% occurred
during the flare-peak interval (elapsed time $t$ = 7 - 13 ks)
but pileup was negligible at   $\leq$5\% after the flare had 
decayed ($t >$ 25 ks). Since the  pileup was moderate and only 
present over a short $\sim$6 ks time interval spanning about
10\% of the exposure time, it did not adversely affect the
data.  The  telltale  ``donut hole'' source image characteristic 
of severe pileup was  not seen and any pileup-induced broadening 
of the source image amounted to no more than a few percent (Sec. 4.3.5).
Nevertheless, we have exercised considerable caution in
our image and spectral analysis to insure that our results 
are not affected by the moderate pileup.

\subsection{RY Tau Images and Extended X-ray Structure}

\subsubsection{X-ray Positions}

The broad-band (0.2 - 8 keV)  X-ray centroid position 
of RY Tau measured with the HEASOFT XIMAGE {\em centroid}
\footnote{http://heasarc.gsfc.nasa.gov/docs/xanadu/ximage/ximage.html}
tool using source counts centered inside a box of half-width
3$''$ is J042157.41$+$282635.24. This result is unchanged if
the box half-width is reduced to 1.$''$5. The CIAO {\em dmstat} 
tool gives an identical centroid position.  The X-ray centroid is offset
0.$''$24 south of  the {\em HST} GSC vers. 2.3.2 position of
RY Tau and 0.$''$33 southwest of the 2MASS position 
(Table 1 notes; Fig. 3). These small offsets are  within
{\em Chandra}'s ACIS-S absolute astrometric accuracy of 
$\approx$0.$''$42 (90\% confidence) for on-axis 
sources\footnote{http://asc.harvard.edu/proposer/POG/ }.

Since the flare resulted 
in a flux increase at higher energies above 2 keV (Sec. 4.4)
we compared X-ray centroids in soft (0.2 - 2 keV) and 
hard-band (2 - 8 keV) images. Any significant offset 
would obviously be of interest since it could signal
a second star in the system. To avoid any possible effects 
of pileup, we measured the positions using events recorded
during the initial flare rise segment   ($t$ = 0 - 7 ks) and the 
post-flare segment ($t$ $>$ 25 ks). A small offset is 
seen in both time segments but it is of low significance.
Specifically, the hard-band centroid is offset to the northwest 
of the soft-band centroid by $\approx$0.$''$03 - 0.$''$05 along 
P.A. $\approx$300$^{\circ}$ - 320$^{\circ}$. The larger offset
corresponds to the flare rise phase.   Although the offset is too 
small to be considered statistically significant, the P.A.
is consistent with that of the putative {\em Hipparcos}
companion (Sec. 2) and again raises the question of whether a
second object might  be present.

\subsubsection{Azimuthal  Count Distribution}
Evidence  for X-ray extension toward the  
northwest is seen when extracting  counts 
within non-intersecting  regions of equal angular and radial size  
distributed along different directions from the X-ray source.
The wedge-shaped regions used for this count extraction 
subtend an angular width of
10$^{\circ}$ and are restricted to radii within
the range 1$''$ $\leq$ $r$ $\leq$ 3$''$ from the
X-ray peak. As Figure 3 shows, a significant excess in 
soft-band counts  (0.2 - 2 keV) is present along
the optical jet axis at P.A. $\approx$ 295$^{\circ}$
and also at higher P.A. $\approx$ 310$^{\circ}$ - 330$^{\circ}$.
The excess is present in images based on the full exposure
and in images restricted to post-flare data only ($t$ $>$ 25 ks).
It is also present along P.A. $\approx$ 305$^{\circ}$ - 325$^{\circ}$ 
in images with sub-pixel event repositioning 
(SER)\footnote{http://cxc.harvard.edu/ciao/why/acissubpix.html} applied.
No significant excess was found in any of the other three
quadrants. A similar analysis based on medium-energy 
2 - 4 keV images shows  a weak excess toward the northwest,
but it is of low statistical significance and its
morphology is not stable as a function of the number of 
iterations in deconvolved images  (Sec. 4.3.4). Thus,
a deeper exposure will be needed to determine whether any 
significant excess is present at energies above 2 keV.

\subsubsection{PSF Asymmetry}
The regions used to extract the source counts plotted in
Figure 3 exclude  counts at small offsets of $r$ $<$ 1$''$.
This precaution was necessary because a
recently-discovered asymmetry in the {\em Chandra}
PSF can produce artificial structure at radii 
0.$''$6 $<$ $r$ $<$ 1$''$ over a limited range of position angles 
P.A.$_{asymm}$ = 195$^{\circ}$ $-$ ROLL ($\pm$25$^{\circ}$).
For the RY Tau observation, the nominal roll angle was
ROLL = 268.6$^{\circ}$ so 
P.A.$_{asymm}$ = 286.4$^{\circ}$ ($\pm$25$^{\circ}$).
The region that may be affected by the PSF asymmetry is
plotted in Figure 4. The asymmetry produces hook-shaped
artificial structure within the central arcsecond in {\em Chandra} 
HRC images and there is evidence that the asymmetry is also
present in ACIS data. There is no indication that the 
asymmetry affects image structure at radii $r$ $>$ 1$''$.
The asymmetry was reported in October
2010 after the RY Tau  observation was obtained
\footnote{Current information on the PSF asymmetry  can be found at:~ \\
http://cxc.harvard.edu/ciao/caveats/psf\_artifact.html}.

\subsubsection{Image Deconvolution}
Since the  initial image analysis (Sec. 4.3.2) suggested that extension
toward the northwest is present in the soft-band, we constructed several 
different deconvolved images  using CIAO {\em arestore}, which is 
based on the Lucy-Richardson method (Lucy 1974; Richardson 1972).
This procedure removes some of the blurring effect on the source 
image due to the telescope optics. We obtained deconvolved images 
in several different energy bands but focus here on images 
in the  soft 0.2 - 2 keV  band. We  compared deconvolved images 
generated using events from the full exposure with those using only events from
the low-pileup post-flare segment. The
deconvolution used observation-specific energy-filtered 
PSF image files created using the 
{\em Chart} and {\em MARX} simulators according to CIAO
science thread procedures.  These PSF images
take into account the source spectrum during the time
interval of interest  and the source position 
on the CCD relative to the optical axis.

Extended structure toward the northwest is clearly
seen in deconolved images in the 0.2 - 2 keV band.
The extension is present in deconvolved images generated 
with and without pixel randomization, and is visible in
both full-exposure and post-flare ($t$ $>$ 25 ks) images.
Figure 4 shows the raw 0.2 - 2 keV image before deconvolution,
along with deconvolved images for the full-exposure and
post-flare time interval. For comparison, deconvolved 
images using 50 and 100 iterations in {\em arestore} are shown. 

The morphology of the extended structure in all of the soft-band
deconovolved images in Figure 4 is quite similar. Overall, the 
extended structure is fainter in the post-flare images since
they are based on only about half the number of soft-band counts 
as the full-exposure images. Also, some variation is expected
since the PSF used in the deconvolution is spectrum-dependent
and the source spectrum was  softer during the post-flare 
segment (Sec. 4.4).   As can be seen, increasing the number of iterations 
from 50 to 100 causes some of the fainter extension to disappear 
and the ``bridge'' connecting the extended structure to the 
star becomes more tenuous. Although some clumpiness is present
in the extended emission, this should not be construed as 
real physical substructure.  Some clumpiness is
expected in  0.$''$125 subpixel images without randomization
(as compared to the physical pixel size of 0.$''$492)
and because of PSF substructure
\footnote{http://cxc.cfa.harvard.edu/cal/Hrma/users\_guide/hrma-notes.pdf}.
The deconvolved images in Figure 4 are based on input images 
without pixel randomization. If pixel randomization is retained,
the extended structure is similar, but slightly  broader.
This is to be expected since randomization broadens the 
source image on the CCD.

As Figure 4 shows, there is some overlap of the extended 
X-ray emission with the region where the known PSF asymmetry
can produce artificial structure. Thus, some of the extended
structure within 1$''$ of the star may be PSF-induced. But,
it is clear from the figure that there is considerable  extended structure
outside of the region affected by the PSF asymmetry.
Specifically, the full-exposure soft-band deconvolved image contains
60  counts at $r$ $>$ 0.$''$6, half of which lie  
outside the PSF asymmetry region. The observed structure
extends outward to at least 1.$''$7 from the X-ray peak, 
well beyond the limit $r$ = 1$''$ where the PSF asymmetry
can affect the image. Furthermore, the extended emission
does not display the characteristic hook-like shape induced
by the PSF asymmetry. We thus conclude that most
of the extended structure is real, but some  contamination 
from the PSF asymmetry may be present at radii $r$ $<$ 1$''$
near P.A. = 286$^{\circ}$.

The extended X-ray structure in the soft-band deconvolved images 
is visible across a range of position angles 
P.A. $\approx$ 285$^{\circ}$ - 325$^{\circ}$. A linear fit 
which takes  the distribution of counts across this  P.A. range
into account gives a best-fit direction 
P.A. = 305$^{\circ}$ for the extension axis, measured relative
to the soft-band X-ray centroid. The extension is visible 
outward to a separation of $\approx$1.$''$7 ($\approx$228 AU)
and inward to  a separation of $\approx$0.$''$6 ($\approx$80 AU).
The extended  structure partially overlaps the inner region of
the optical jet (Fig. 4). The innermost H$\alpha$ jet knot
designated HaA by SB08 lies at an  offset of 1.$''$5 from
the star along P.A. = 299$^{\circ}$, and the knot detected in
[O I] by AA09 lies at an offset of 1.$''$35 along
P.A. = 294$^{\circ}$. But, as Figure 4 shows,  the
soft-band X-ray extension is less-collimated than
the optical jet and extends to  higher P.A. values.
The best-fit soft-band X-ray extension axis at P.A. = 305$^{\circ}$ is in
good agreement with the value inferred for the direction of
the putative companion from {\em Hipparcos} variations (Sec. 2).
It is thus not clear that all of the wide X-ray extension is 
associated with the more tightly collimated optical jet.

\subsubsection{Checks for Flare-Induced PSF  Effects}
Photon pileup can artificially broaden the PSF
\footnote{An example of pileup-induced source broadening
due to a flare can be found at:~
http://cxc.cfa.harvard.edu/cal/ASPECT/psf\_degrade/}.
As  mentioned above (Sec. 4.2), moderate pileup of 10\% - 13\% 
occurred during the flare peak time interval, 
It is thus important to note that the extended X-ray structure
is still present in deconvolved images based on low-pileup
post-flare data (Fig. 4). This provides added confidence
that the X-ray extension is not due to pileup-induced
PSF broadening.

We checked for source broadening by applying  Gaussian fits
to the source image during the flare-peak time interval  ($t$ = 7 - 13 ks)
and the post-flare phase ($t$ $>$ 25 ks). The source profile
was fitted separately in the north-south and east-west directions.
No energy filtering was applied. We found no evidence of 
broadening of the source FWHM during the flare at levels 
above $\approx$3\%.

As a further check for changes in the source size during the 
flare, we plotted the event  positions in (X,Y) sky-pixel 
coordinates as a function of time (Fig. 5). The (X,Y) values
correspond to (R.A., Dec.) but are in units of pixel coordinates.
As Figure 5 shows, there is more scatter in (X,Y) during the
flare than afterward. Using (X,Y) data binned at 5 ks intervals,
we find that the standard deviation of the mean (X,Y) positions
for each bin is about 3\% - 6\% larger during the flare than
afterword. This is comparable to, but slightly larger than that
inferred above from Gaussian fits. Some of this difference is
likely due to a low-level of flare-induced PSF broadening but
a flaring close companion could also cause positional scatter.

\subsection{The X-ray Spectrum}

The ACIS-S CCD spectrum shown in Figure 6 was extracted
from a circular region of radius 3$''$ centered on
the star  and thus  includes emission from both
the star and the extended X-ray structure.
No spectrum of the jet-like feature itself was obtained due to 
its faintness and proximity to the bright stellar source.
Several emission lines, or line blends, are visible
in the spectrum. These lines reflect a broad range of 
X-ray temperatures. Cool plasma is revealed by the
O VII He-like triplet  (E$_{lab}$ = 0.57 keV) which
forms at a characteristic temperature T $\sim$ 2 MK
and the Ne IX triplet (E$_{lab}$ = 0.92 keV) which
forms at T $\sim$ 4 MK. At the other extreme, we also
detect the Fe K$\alpha$ complex which forms in very
hot plasma at T $\sim$ 40 MK.

We extracted separate spectra for the flare rise  
($t$ = 0 - 7 ks; 2350 counts), flare decay ($t$ = 13 - 30 ks; 3848
counts), and late post-flare ($t$ = 30 - 58.6 ks; 4011 counts) segments.
The above segments excluded the time interval
$t$ = 7 - 13 ks near flare peak during which photon
pileup reached its maximum of $\approx$10\% - 13\%
and could result in some artificial hardening of
the energy spectrum.
Each spectrum was fitted with an absorbed two-temperature (2T) $vapec$
optically thin plasma model consisting of  a cool
(kT$_{1}$) and hot (kT$_{2}$) component. Extremely hot 
plasma at a temperature kT$_{2,flare}$ $\gtsimeq$ 9 keV (T $\gtsimeq$ 100 MK)
was present during the flare rise and decay segments,
and the Fe K$\alpha$ line complex was strongest during the flare 
rise segment (Fig. 6-middle). During the post-flare segment, the temperature 
of the hot plasma component had decreased by a factor of $\sim$2 and the 
count rate was more stable. But, high-temperature lines are
still  present in the post-flare spectrum including faint 
Fe K$\alpha$ emission and strong emission from the Si XIII
He-like triplet at 1.86 keV.  The late post-flare spectrum has
negligible pileup ($<$5\%) and its best-fit parameters
are given in Table 2. They are similar to values found for
other TTS.  The best-fit absorption N$_{\rm H}$ = 5.5 [4.7 - 6.8]
$\times$ 10$^{21}$ cm$^{-2}$ equates to 
A$_{\rm V}$ = 2.5 [2.1 - 3.1] mag (Gorenstein 1975 conversion) 
which is consistent with the optical value 
A$_{\rm V}$ = 2.2 $\pm$ 0.2 mag (Calvet et al. 2004). 
The unabsorbed X-ray luminosity during 
the post-flare segment was log L$_{\rm X}$(0.2 - 10 keV) = 30.7 ergs s$^{-1}$,
but was at least 0.7 dex larger during the flare. 

Although the 2T $vapec$ model generally produces an acceptable fit of the 
post-flare spectrum, it underestimates the flux in the weak emission 
feature near 0.57 keV that is likely the O VII He-like triplet (Fig. 6-bottom).
This flux deficit is still present even if the O abundance is allowed to
vary in the fit. This suggests that the O VII line may be of nonstellar
origin, arising in very cool plasma in the jet or an accretion shock. 
Adding a Gaussian 
component to the 2T $vapec$ model at a fixed width FWHM = 110 eV
corresponding to ACIS-S spectral resolution improves the 
fit slightly and gives an unabsorbed continuum-subtracted line flux
F$_{\rm X,O VII}$ = 4.7 $\times$ 10$^{-15}$ ergs cm$^{-2}$ s$^{-1}$.

\vspace*{0.5in}

\section{Discussion}

\subsection{X-ray Emission from the Jet?}

The extended  structure shown in Figure 4 overlaps the 
inner region of the optical jet. This raises the interesting
possibility that the extended X-ray emission arises in the jet, close
to the star. But, as discussed below,
there are legitimate questions as to whether the RY Tau jet
speed and shock speed are sufficient to produce soft-band thermal  
X-rays by shock-heating alone.

The predicted temperature for a shock-heated jet
with a shock speed $v_{s}$ is 
(Raga et al. 2002):

\begin{equation}
T_{s} = 1.5 \times 10^{5}\left[\frac{v_{s}}{100~\rm{km~s^{-1}}}\right]^{2}~K.
\end{equation} 
Assuming the jet impacts a stationary target
($v_{s}$ $\approx$ $v_{jet}$ $\approx$ 165 km s$^{-1}$),
then the maximum shock temperature
is $T_{s}$ $\approx$ 0.4 MK (k$T_{s}$ $\approx$ 0.035 keV).
This temperature is a factor of $\sim$3 lower than 
that needed to produce detectable thermal X-ray emission 
at  kT $\approx$ 0.1 - 0.2 keV (T $\approx$ 1 - 2 MK),
below which {\em Chandra} has little sensitivity.
The detection of soft-band X-rays from the jet thus seems to 
require  either a higher jet speed (and shock speed) than assumed 
above, or an additional heating mechanism other than shocks.
The deprojected terminal speed of the jet is uncertain by a 
factor of $\sim$2 (AA09). Doubling
the shock speed in the above calculation to
$v_{s}$ = 330 km s$^{-1}$ would increase the shock 
temperature to 
$T_{s}$ $\approx$ 1.6 MK (k$T_{s}$ $\approx$ 0.14 keV),
which is high enough to produce soft X-ray emission.
But, a shock speed of $v_{s}$ = $v_{jet}$ $\approx$ 330 km s$^{-1}$
is at odds with optical data, which suggest 
a {\em lower} shock speed in the range 
$v_{s}$ $\approx$ 20 - 50 km s$^{-1}$ (AA09).
Unless very high velocity jet plasma has escaped optical 
detection, it is not obvious that the jet speed is sufficient
to produce shock-heated plasma at the X-ray temperatures
sufficient for detection  in the {\em Chandra} soft band.

If the soft-band extended X-ray emission detected by {\em Chandra} 
is due to the jet, then the above estimates 
suggest that other mechanisms besides shocks are needed to 
heat the jet to  X-ray emitting temperatures. In this regard, 
magnetic fields may  play a role. Recent
observations have in fact given strong support for the 
presence of magnetic fields in YSO jets.
A notable discovery
is the detection of linearly polarized radio emission
in the HH 80-81 jet (Carrasco-Gonz\'{a}lez et al. 2010).
Linearly polarized radio emission is
typical  of synchrotron radiation and 
is produced by relativistic particles trapped in a
magnetic field threading the HH 80-81 jet. The
new results for HH 80-81 raise the interesting 
question of whether the RY Tau jet might also show 
signs of nonthermal radio emission that could signal 
an entrained magnetic field. 

Even if the jet is contributing to the extended soft-band
X-ray emission, its contribution to
the total X-ray luminosity of RY Tau is predicted to
be very low. The predicted
intrinsic (unabsorbed) luminosity of the shocked jet is 
the lesser of the two values
corresponding to radiative and non-radiative cases (eqs. 8 and 9
of Raga et al. 2002).
For the range of parameters relevant to RY Tau, the minimum corresponds
to the radiative shock case: 

\begin{equation}
L_{r,jet} = 4.1 \times 10^{-6} \frac{n_{o}}{100~ \rm{cm^{-3}}}\left[\frac{r_{bs}}{10^{16}~ \rm{cm}}\right]^{2}\left[\frac{v_{s}}{100~ \rm{km~ s^{-1}}}\right]^{5.5} L_{\odot}
\end{equation}

where $n_{o}$ is the preshock number density, 
$r_{bs}$ is the characteristic radius of the bow shock around 
its axis,  and $v_{s}$ is the shock speed.
For purposes of an estimate, we adopt the following values used by
AA09. The electron density in the
RY Tau jet is not well-determined but is assumed to be similar 
to that of other YSO jets studied at high angular resolution 
with a value that decreases with radius from the star
$n_{o}$ = 5000(1$''$/$r$) cm$^{-3}$ ($r$ $\geq$ 0.$''$5). 
For the range of offsets at which the X-ray extension is 
visible (Fig. 3), the value $n_{o}$ 
= 5000 cm$^{-3}$ at $r$ = 1$''$ is a reasonable estimate.
The optical jet full-width is nearly constant as a function of
distance from the star (AA09) and we adopt  a full-width of 0.$''$3,
which equates to a jet radius $r_{bs}$ = 20 AU = 
3 $\times$ 10$^{14}$ cm (d = 134 pc).
Assuming a  shock speed $v_{s}$ = $v_{jet}$ = 165 km s$^{-1}$ (SB08)
gives $L_{r,jet}$ = 1.1 $\times$ 10$^{28}$ ergs s$^{-1}$. It
is obvious from equation (2) that the  above  estimate is 
quite sensitive to the assumed shock speed $v_{s}$.

Could soft X-ray emission from a shocked jet of the above  luminosity
$L_{r,jet}$ = 1.1 $\times$ 10$^{28}$ ergs s$^{-1}$ generate
sufficient counts to explain the extended X-ray structure?
To address this question, we assume that the jet has a
simple thermal spectrum with a characteristic shock temperature
kT$_{s}$ = 0.035 keV corresponding to $v_{s}$ = $v_{jet}$ = 
165 km s$^{-1}$ (eq. 1). The absorption column density
toward the extended jet-like structure  is not directly known from
observations, but since the structure  is offset from the star
we assume an extinction toward the jet of A$_{\rm V}$ = 1 mag,
or about half the stellar value  (Sec. 2). This extinction
corresponds to an absorption column density 
N$_{\rm H}$ = 1.6 $\times$ 10$^{21}$ cm$^{-1}$ (Vuong et al. 2003).
Using this simple model and an unabsorbed flux density
F$_{\rm X,unabsorbed}$ = 5.1 $\times$ 10$^{-15}$ ergs cm$^{-2}$ s$^{-1}$
corresponding to the above value of $L_{r,jet}$ (d = 134 pc), the
{\em Chandra} PIMMS simulator predicts that $<$1
count will be detected in the 0.2 - 2 keV range in a 56 ks
ACIS-S exposure. By comparison, the extended structure
in the soft-band deconvolved image contains 30 counts, 
excluding those inside the PSF asymmetry region (Sec. 4.3.4).

In order to attribute  the 30 soft-band extended counts 
lying outside the PSF asymmetry region 
to a shocked jet, a higher jet luminosity 
$L_{r,jet}$ $\approx$ 1.2 $\times$ 10$^{29}$ ergs s$^{-1}$  
is needed. This value is one-seventh  of the 
soft-band luminosity (star $+$ extension) of 
L$_{X,1}$ = 8.4 $\times$ 10$^{29}$ ergs s$^{-1}$
determined from spectral fits (Table 2). To achieve
this higher jet luminosity, the values of one or more
of the three jet parameters in equation (2) would need to
be increased. If the values of $n_{0}$ and
$r_{bs}$ are held fixed to those  used above, then 
a higher shock speed $v_{s}$ $\approx$ 254 km s$^{-1}$
(kT$_{s}$ $\approx$ 0.084 keV) would be required. Although
this shock speed is plausible given that the deprojected
jet speed is uncertain by a factor of $\sim$2, it is
at least five times greater than shock speeds inferred
from optical data (AA09).

\subsection{Plasmoid Ejections?}

As shown above, it is difficult to account for shock-induced
X-ray emission in the jet based on the  optically-determined
values $v_{jet}$ = 165 km s$^{-1}$ and shock speeds 
$v_{s}$ $<$ 100 km s$^{-1}$. If the extended soft-band X-ray
emission originates in the jet, then either higher speeds
are needed or other jet heating mechanisms besides shocks
are required.

Plasmoid ejections provide one possible means of attaining
higher speeds and sufficiently high plasma temperatures  
to produce X-ray emission   The   model developed by
Hayashi et al. (1996) predicts that hot plasmoids 
(T$_{p}$ $\sim$ 10$^{7}$ - 10$^{8}$ K) will
be ejected at high speeds in bipolar directions during hard 
protostellar X-ray flares. As the plasmoids move outward
and cool, they may be revealed optically as inhomogeneities
within the jet. The plasmoid ejection speed is predicted to be 
$v_{p}$ $\sim$ (2 - 5) $\times$ $v_{K}$,
where $v_{K}$ = $\sqrt{GM_{*}/R_{in}}$ is the Keplerian rotation 
speed at the inner edge of the disk. For RY Tau
($R_{in}$ = 0.3 AU, M$_{*}$ =  1.7 M$_{\odot}$; 
Schegerer et al. 2008) we obtain $v_{K}$ = 71 km s$^{-1}$ and 
plasmoid speeds   $v_{p}$ $\sim$ 140 - 350 km s$^{-1}$. 
Hot ejected plasmoids moving outward
at speeds of several  hundred km s$^{-1}$ through the stellar 
or disk wind, or plowing into accreting gas, would be capable
of producing soft X-rays from shocks. Specifically,
adopting $v_{s}$ = 350 km s$^{-1}$  gives T$_{s}$ = 1.8 MK
or kT$_{s}$ = 0.16 keV) (eq. 1), which lies near the
low end of the temperature range detectable by 
{\em Chandra} ACIS-S.

In order for hot plasmoids to escape, they would need to
be ejected at (or accelerated to) speeds in excess of
the escape speed. At the surface of RY Tau, the escape speed 
is $v_{esc}$ =  410 km s$^{-1}$,
assuming M$_{*}$ = 1.7 M$_{\odot}$ and 
R$_{*}$ =  3.85 R$_{\odot}$ (Schegerer et al. 2008).
This value of $v_{esc}$ is  slightly above the maximum 
predicted plasmoid ejection speed in the Hayashi et al. 
model.  But, the plasmoids are expected to form in the 
flaring region above the stellar surface where the 
escape speed will be less. A plasmoid that has escaped
at an  ejection speed  $v_{p}$ $\sim$ 350 km s$^{-1}$ 
would traverse a projected distance equivalent to 1$''$
in 1.8 years at the distance to RY Tau. However,  this
assumed speed could  be an underestimate given that coronal
mass ejections (CMEs)  from the Sun can reach speeds of 
$\geq$1000 km s$^{-1}$ (Aarnio et al. 2011; Gopalswamy et al. 2005).
The Hayashi et al. model predicts that a cold dense disk 
wind will form. Ejected plasmoids moving outward would
create disturbances or shocks in the disk wind (or the
stellar wind). By analogy, interactions of CMEs with the 
solar wind are indeed known to occur and can produce shocks and
wind disturbances  (Gosling et al. 1995; Wang et al. 2001).   
Plasmoids that fail to reach
escape speed could also shock against the outflowing wind on
reentry. Variations on the plasmoid model have
been proposed to explain hot X-ray 
plasma detected near OB stars (Howk et al. 2000;
Waldron \& Cassinelli 2009).

The above discussion focuses on shock emission at
temperatures of a few MK that could be produced by 
high-speed plasmoids.  But, if the plasmoids
have temperatures T$_{p}$ $\sim$ 10$^{7}$ - 10$^{8}$ K
then they could themselves be sources of harder
X-rays (kT$_{p}$ $\sim$ 1 - 8 keV), independent of any 
soft shock-induced emission. 
As a representative 
case, we consider an ejected  plasmoid of temperature
T$_{p}$ $\sim$ 10$^{7}$ - 10$^{8}$ K and number density 
n$_{p}$. The radiative cooling time of the plasmoid is
$\tau_{rad}$ = 3kT$_{p}$/n$_{p}$$\Lambda$(T$_{p}$),
where  $\Lambda$(T) is the plasma emissivity
per unit emission measure. As Figure 10 of 
Audard et al. (2004) shows, the solar-abundance
value of   $\Lambda$(T) over the temperature range
10$^{7}$ -  10$^{8}$ K is nearly constant with
an average value  $\Lambda$ $\sim$ 
2.5 $\times$ 10$^{-23}$ ergs cm$^{3}$ s$^{-1}$.
Using this average value for $\Lambda$, the radiative
cooling time becomes 
$\tau_{rad}$ $\sim$ (5 - 50)/n$_{p,6}$ years, 
where n$_{p,6}$ is the plasmoid density in 
units of 10$^{6}$ cm$^{-3}$. As noted above,
it would take about 2 years for a high-speed
plasmoid to move out to a separation of 
1$''$ where the extended X-ray structure is 
seen. If we require $\tau_{rad}$ $\gtsimeq$ 2 y
then the plasmoid density cannot be much
larger than n$_{p}$ $\sim$ 10$^{6}$ - 10$^{7}$
cm$^{-3}$.  

The above density constraint only takes  radiative
cooling into account and is thus  a very rough estimate.
More detailed numerical simulations would 
be needed to determine how the temperature, density
and speed  of an ejected plasmoid evolve as it 
moves away from the T Tauri star and interacts
with surrounding material. Such simulations 
would need to take into account such potentially
important factors as adiabatic cooling,  the 
influence of any jet magnetic field on plasmoid
expansion, and plasma instabilities. Obviously,
the physical picture could be considerably more
complicated for a TTS than for the Sun. If it can be demonstrated
that hot ejected plasmoids can survive for a few years
(or less, if they are moving at the higher speeds observed
for some solar CMEs),
then a trail of such  plasmoids ejected during repetitive
flares could be revealed as  high-temperature
plasma offset from the star, perhaps explaining
the X-ray  extension seen in the RY Tau images.

\subsection{A  Close Companion?}
As previously noted (Sec. 4.3.4), the soft-band X-ray
extension is centered along an axis toward  
P.A. $\approx$  305$^{\circ}$ from the X-ray peak,
but the extension spans a range of $\pm$20$^{\circ}$
in position angle. Interestingly, the central axis
direction is nearly identical to that inferred for
the putative companion from {\em Hipparcos} variations (Sec. 2).
But, near-IR speckle observations by Leinert et al. (1993) probed
the separation range 0.$''$13 - 13$''$ and found no companion.

The only objects reported so far at  separations similar to
the range $\approx$0.$''$6 - 1.$''$7 of the X-ray extension 
are the H$\alpha$ knot HaA at a separation of 1.$''$5 along
P.A. = 299$^{\circ}$ (epoch 2005.15; SB08) and the
[O I] (6300~ \AA) knot at an offset of 1.$''$35 from the
optical centroid along P.A. = 294$^{\circ}$ (epoch 2002.04; AA09).
These knots are believed to be non-stellar shocked structures,
and this interpretation is supported by the absence of
continuum emission at the [O I] knot position (AA09).
The above knot separations were measured in optical 
observations taken 5 - 8 years prior to our {\em Chandra}
observation and both knots could now have moved outward
to separations of $>$2$''$ (SB08). In that case, it 
would seem unlikely that the X-ray extension is related
to these knots specifically, but more recently ejected knots could
be present at separations of $<$1.$''$7 where the X-ray
extension is seen.

Given that the near-IR speckle observations of Leinert et al. (1993)
found no companion within the offset range 0.$''$13 - 13$''$, it
would appear that any companion  must lie very close to RY Tau.
But, searches for a companion at very close separations have 
also yielded negative results. Keck Interferometer observations
at 2 $\mu$m found no evidence for a companion within
the separation range 2.5 - 30 mas (Pott et al. 2010). 
The VLTI observations obtained by Schegerer et al. (2008) 
probed separations down to $\sim$1 AU ($\sim$75 mas) and
neither confirmed nor disproved the existence of the companion 
inferred from {\em Hipparcos} data. 

When limits on {\em Chandra}'s
spatial resolution are considered, it is apparent that any
second X-ray source at a separation  of $<$0.$''$4 from RY Tau
would not be fully resolved.
The on-axis ACIS-S PSF core has FWHM = 0.$''$74 at 1.5 keV, and increases 
only slightly with energy below 4 keV, but more steeply above 4 keV 
\footnote{http://cxc.cfa.harvard.edu/cal/Hrma/users\_guide/hrma-notes.pdf}. 
This FWHM value is in good agreement with the simulated PSF in 
Figure 4 which has a half-maximum radius of  $\approx$0.$''$4,
and with the core regions of the soft-band deconvolved images, which are of
similar radius. The only indication from the {\em Chandra} observation 
that a second object might be present at a separation less than the
0.$''$13 speckle imaging lower limit  is the small
$\approx$0.$''$03 - 0.$''$05 offset between the soft and hard
X-ray centroids along 
P.A $\approx$300$^{\circ}$ - 320$^{\circ}$ (Sec. 4.3.1). 
If two X-ray sources are indeed present then there is some ambiguity as to
which one is RY Tau because of the small separation. But,
the absorption in the post-flare X-ray spectrum is consistent
with that expected for RY Tau (Sec. 4.4), so it is reasonable to
assume that the cooler post-flare emission is due to RY Tau.
In that case, one would identify the soft-band X-ray peak with
RY Tau and the harder peak to the northwest with the (flaring) putative
companion. But we caution that the offset between the soft and hard
centroids, although suggestive,
is of low significance and 
smaller than what can be reliably measured with {\em Chandra}.
Clearly, any definite proof that a second object
is present at a small offset northwest of RY Tau 
will need to come from higher spatial resolution 
observations. Infrared or radio interferometry capable of 
detecting a more heavily-obscured object (possibly a 
flaring protostar) would seem to offer the most promise.

%%Spectral fits during the flare-rise segment
%%indicate that the hot flaring plasma was seen through an absorption
%%column N$_{\rm H}$ $\approx$ 1.5 [1.4 - 1.7] $\times$ 10$^{22}$ cm$^{-2}$.
%%This absorption corresponds to A$_{\rm V}$ = 6.8 [6.3 - 7.7] mag
%%(Gorenstein 1975), and is about 3 times larger than that 
%%inferred from the post-flare spectrum (Table 2). Thus, if
%%a second object was indeed responsible for the flare then it is
%%more heavily absorbed than RY Tau,  raising  the obvious
%%question of whether it might also be driving the jet.

\subsection{Low-Temperature Emission Lines}

The origin of the low-temperature O VII and Ne IX
emission lines in the RY Tau spectrum is of 
considerable interest. Although cool plasma in the
jet may contribute to this line emission, an
accretion shock origin is not yet ruled out.
In this regard, we note that these two lines are present
even in a spectrum extracted from a small circular region
of radius 0.$''$5  (= 1 ACIS physical pixel) centered on the 
star. This region lies closer to the star than the
extended  structure shown in Figure 4.
In order to further constrain the origin of these
low-temperature lines, X-ray grating spectra which can 
separate the He-like triplet components and provide electron
density information will be required. Such grating
spectra have been obtained  for the accreting TTS
TW Hya and  high densities have been inferred 
from the O VII line, suggesting that it forms in 
the accretion shock or postshock region 
(Kastner et al. 2002;  Brickhouse et al. 2010).

\section{Summary}

A sensitive {\em Chandra}  observation of  RY Tau reveals
complex X-ray emission including a  cool steady plasma 
component, a variable superhot component with a maximum
flare temperature T $\gtsimeq$ 100 MK, and faint structure
extending outward from the star toward the northwest. 
The flaring hot component is no doubt of magnetic origin but 
a small (low-significance) offset between the soft and hard 
X-ray centroids raises the intriguing question of whether the flare
occurred on RY Tau or a second object.
In contrast, some of the coolest plasma (including the faint
O VII emission line) likely arises in the jet or an 
accretion shock. Higher resolution X-ray grating spectra will
be needed to place tighter constraints on the electron density 
in the cool plasma in order to  distinguish between jet or accretion
shocks and any cool coronal component.

Deconvolved soft-band (0.2 - 2 keV) images show faint structure 
extending outward to a separation of $\approx$1.$''$7 from the star. 
The X-ray  extension is visible over a range of position angles
spanning P.A. = 285$^{\circ}$ - 325$^{\circ}$ with a central
axis directed toward P.A. $\approx$  305$^{\circ}$. Most of
the extended structure at offsets of $>$1$''$ is believed to be
real, but some artificial extension may be present at smaller
offsets of $<$1$''$ due to a known asymmetry in the {\em Chandra} 
PSF directed along P.A. $\approx$ 286$^{\circ}$ ($\pm$25$^{\circ}$).

The extended X-ray structure overlaps the blueshifted optical 
jet axis at P.A. $\approx$ 295$^{\circ}$ and some of the 
X-ray extension could thus be jet-related. But, predicted shock 
temperatures based on  optically-determined jet parameters are 
lower than required to achieve X-ray temperatures of a few million K.
It is possible that the hottest jet material is moving faster 
than  optical data suggest and thereby producing hotter 
shocked plasma in the jet than might otherwise be expected.
Higher-temperature plasma could also be present in the 
jet if magnetic heating is important.  Hot high-velocity
plasmoids ejected  during repetitive hard X-ray flares could also 
give rise to extended X-ray emission as a result of their own 
thermal radiation or by shocking onto intervening material.
The plasmoid interpretation is somewhat speculative but is
supported by some theoretical YSO flare models and it is 
well-established that the Sun (a much less active star)
commonly launches CMEs at speeds $>$1000 km s$^{-1}$.
A final possibility is that the X-ray extension is 
related to a close companion to the northwest of RY Tau
whose presence has been inferred from {\em Hipparcos} 
variations. However, sensitive ground-based searches
for a companion have so far yielded negative results.

\acknowledgments

This work was supported by {\em Chandra} award GO0-11028X issued by the 
Chandra X-ray Observatory Center (CXC). The CXC is operated by the 
Smithsonian Astrophysical Observatory (SAO) for, and on behalf of, 
NASA under contract NAS8-03060. MA acknowledges support from  
Swiss National Science Foundation grants PP002-110504 and
PP00P2-130188. We thank V. Kashyap (CXC) for current information on
ACIS PSF structure and independent  PSF analysis of RY Tau.

\clearpage

%%% Comment out  the following line to include Tables
%%\end{document}

% TABLE1.TEX

\clearpage

%% TABLE 1
\begin{deluxetable}{lllllll}
\tabletypesize{\small}
\tablewidth{0pt} 
\tablecaption{X-ray Sources Near RY Tau }
\tablehead{ 
	   \colhead{Object}	        &
           \colhead{R.A.}               &
           \colhead{Decl.}              &
           \colhead{Net Counts}         &
           \colhead{E$_{50}$}           &
           \colhead{Offset}             &
           \colhead{P.A.}            \\
           \colhead{} 	                &
           \colhead{(J2000)}                 &
           \colhead{(J2000)} &
           \colhead{(cts)}                   &
           \colhead{(keV)}                   &               
           \colhead{($''$)}                   &                  
           \colhead{(deg.)} 
                         }                                           
\startdata
RY Tau          & 04 21 57.41\tablenotemark{a} & $+$28 26 35.24\tablenotemark{a} & 12366 $\pm$ 111  & 2.04 & ...  & ... \\
1 		& 04 21 54.97 & $+$28 27 15.53 & 4$\pm$2\tablenotemark{b} & 0.50 & 51.6 & 321.4  \\
2\tablenotemark{c} & 04 21 55.93 & $+$28 26 47.55 & 12$\pm$4 & 2.12 & 23.1   & 302.2  \\
3               & 04 21 58.22 & $+$28 26 19.78 & 43$\pm$7 & 1.82 & 18.8   & 145.3  \\
4               & 04 22 00.09 & $+$28 26 13.39 & 5$\pm$2  & 2.56 & 41.5   & 121.8   \\
5               & 04 22 01.11 &	$+$28 27 19.68 & 4$\pm$2\tablenotemark{b} & 0.44 & 66.0 & 47.7   \\
\enddata
\tablecomments{
X-ray data are from CCD7 (ACIS chip S3) using events in the 0.3 - 8 keV range inside the 3$\sigma$ source
extraction ellipse.   
Tabulated quantities are: J2000.0 X-ray position (R.A., Decl.),  
net counts and net counts error from {\em wavdetect} (accumulated in a 55766 s exposure, rounded 
to the nearest integer,
background subtracted and PSF-corrected); median photon energy (E$_{50}$),
offset and P.A. (measured east from north) of the source, 
relative to the RY Tau X-ray position.
}
\tablenotetext{a}{~The {\em Chandra} position is offset 0.$''$24 south
                   of the {\em HST} GSC vers. 2.3.2 position of RY Tau
                   (HST J042157.408$+$282635.48) and 0.$''$33 southwest
                   of the 2MASS near-IR position (2MASS J042157.401$+$282635.55).}
\tablenotetext{b}{~Low significance detection.}
\tablenotetext{c}{~This X-ray source may be associated with optical emission knot
                   Ha6 listed in Table 2 of SB08. See text (Sec. 4.1).  }
\end{deluxetable}

\clearpage

% TABLE2.TEX 
\begin{deluxetable}{lc}
\tabletypesize{\scriptsize}
%\tablewidth{33pc}
\tablewidth{0pc}
\tablecaption{RY Tau Spectral Fits  
   \label{tbl-1}}
\tablehead{
\colhead{Parameter}      &
\colhead{Value  }
}
\startdata
Model                                   & 2T $vapec$                               \nl
Time Interval                           & post-flare                              \nl
N$_{\rm H}$ (10$^{21}$ cm$^{-2}$)       & 5.5 [4.7 - 6.8]                         \nl
kT$_{1}$ (keV)                          & 0.59 [0.46 - 0.70]                      \nl
norm$_{1}$ (10$^{-3}$)                  & 0.39 [0.20 - 0.88]                      \nl 
kT$_{2}$ (keV)                          & 4.82 [4.20 - 5.67]                      \nl
norm$_{2}$  (10$^{-3}$)                 & 1.14 [1.04 - 1.24]                      \nl
Abundances                              & varied\tablenotemark{a}                 \nl
$\chi^2$/dof                            & 285.0/282                               \nl
$\chi^2_{\nu}$                          & 1.01                                    \nl
F$_{\rm X}$ (10$^{-12}$ ergs cm$^{-2}$ s$^{-1}$)              & 1.27 (2.16)       \nl
F$_{\rm X,1}$ (10$^{-12}$ ergs cm$^{-2}$ s$^{-1}$)            & 0.09 (0.39)       \nl
F$_{\rm X,2}$ (10$^{-12}$ ergs cm$^{-2}$ s$^{-1}$)            & 1.18 (1.77)       \nl
log L$_{\rm X}$ (ergs s$^{-1}$)                               & 30.67             \nl
%%log [L$_{\rm X}$/L$_{bol}$]                                 & $-$5.37           \nl
\enddata
\tablecomments{
Based on  XSPEC (vers. 12.4.0) fits of the background-subtracted ACIS-S spectrum binned 
to a minimum of 5 counts per bin. The fit was restricted to post-flare data
with negligible pileup in the elapsed time range $t$ = 30,000 - 58,627 s (4011 cts). 
The tabulated parameters
are absorption column density (N$_{\rm H}$), plasma energy (kT),
and XSPEC component normalization (norm).
Abundances are referenced to  Anders \& Grevesse (1989).
Square brackets enclose 90\% confidence intervals.
The total X-ray flux (F$_{\rm X}$) and fluxes associated with each model component
(F$_{\rm X,i}$)  are the absorbed values in the 0.2 - 10 keV range, followed in
parentheses by  unabsorbed values. 
The total X-ray luminosity L$_{\rm X}$  is the  unabsorbed 
value in the 0.2 - 10 keV range and  assumes a
distance of 134 pc.}
\tablenotetext{a}{All metal abundances with the exception of Si were fixed at
                  0.25 $\times$ their solar values, based on a best-fit global 
                  metallicity  $Z$ = 0.25 $Z_{\odot}$. The silicon abundance was
                  allowed to vary in order to reproduce the strong Si XIII He-like
                  triplet at  E = 1.86 keV and converged to 
                  Si = 1.6 [1.0 - 2.5] $\times$ solar.}
%%\tablenotetext{a}{Adding a fixed-width Gaussian line at energy E = 565 eV (O VII triplet line)
%%improves the fit and gives $\chi^2$/dof = 230.6/204 ($\chi^2_{\nu}$ = 1.13).}
\end{deluxetable}

\clearpage

%%%% Fig. 1 : FIELD SRC (0.2 - 8 keV base image, 0.5 as pixels)

\begin{figure}
\figurenum{1}
\epsscale{1.0}
\includegraphics*[width=11.0cm,angle=0]{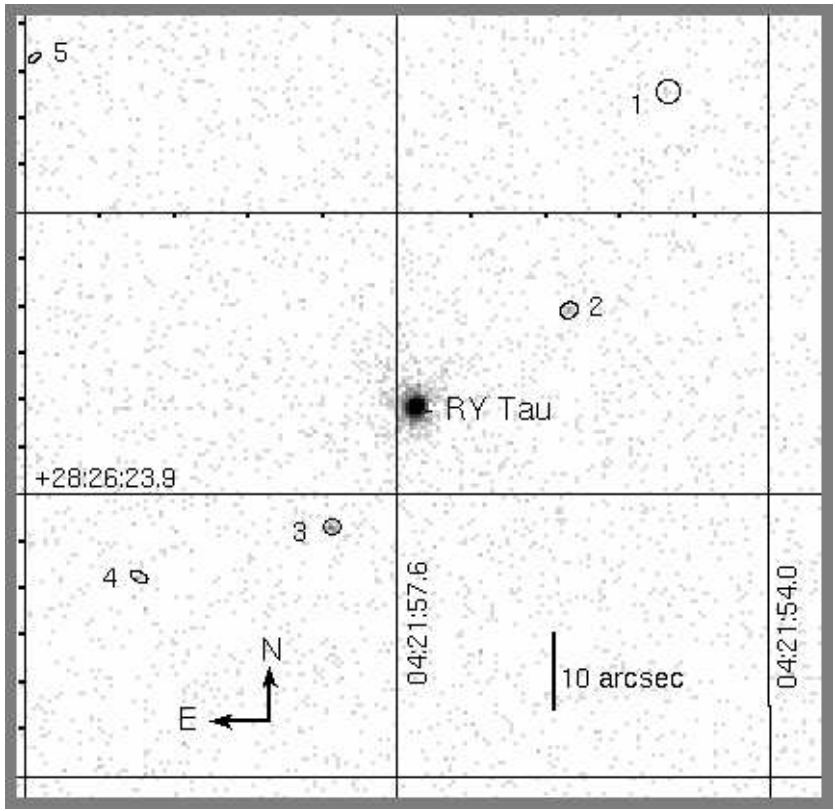} 
\caption{Chandra ACIS-S broad-band (0.2 - 8 keV)  image
of the field near RY Tau. A readout streak running
east-west through the center of the image has been 
removed and filled in with adjacent background.
Source numbers refer to Table 1.
The X-ray centroid position of RY Tau is 
J042157.41$+$282635.24. Pixel size = 0.$''$492. Log intensity scale.}
\end{figure}
\clearpage

%% -------------------------------------------------------
%%%% Fig. 2 Light Curve%%%

\begin{figure}[h]
\figurenum{2}
\epsscale{1.0}
\includegraphics*[height=7.7cm,angle=-90]{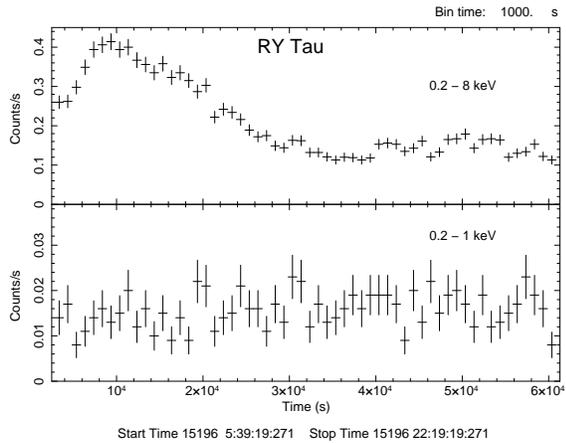}
\caption{Chandra ACIS-S light curves  of RY Tau in 
broad (0.2 - 8 keV) and very-soft (0.2 - 1 keV)
bands. The light curves are based on events extracted
from a cicular region of radius 3$''$ centered on
the source. The broad-band count rate is slightly 
underestimated during the flare-peak interval 
($t$ = 7000 - 13000 s) due to pileup of 
$\approx$10\% - 13\%.  Clear variability is
seen  in the broad-band light curve but 
little or no variability is present in the 
very-soft emission. The binsize is 1000 s. 
Error bars are 1$\sigma$.
}
\end{figure}
\clearpage

%% -------------------------------------------------------                                                       \

%%%% Fig. 3  Counts vs. PA %%%                                                                                       \

\begin{figure}[h]
\figurenum{3}
\epsscale{1.0}
\includegraphics*[height=7.7cm,angle=-90]{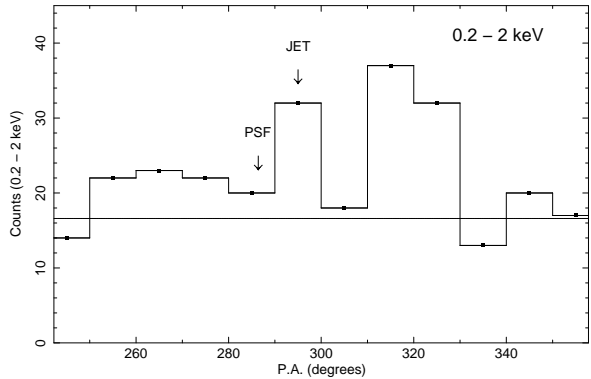}
\caption{Distribution of counts in twelve non-intersecting wedge-shaped
regions of equal size near RY Tau as a function of P.A. (measured east from north). The
counts were extracted from energy-filtered images  (E = 0.2 - 2 keV; no SER or 
deconvolution applied), using the full exposure time (55.766 ks).  Each
region has an angular width  of  10$^{\circ}$ and is restricted to  a radial range
1$''$ $\leq$ $r$ $\leq$ 3$''$ from the X-ray centroid.  The  regions
are centered at P.A. = 245$^{\circ}$, 255$^{\circ}$, ... , 355$^{\circ}$.
The solid horizontal line shows the mean number of counts (mean $\mu$ = 16.6, 
standard deviation $\sigma$ = 3.5) 
computed  from all  regions of the above size encircling the star. The downward arrows mark the
mean  P.A. = 295$^{\circ}$ of the optical jet and the nominal direction of
the known PSF asymmetry at P.A. = 286.4$^{\circ}$. The PSF asymmetry only  produces
artificial extension  at radii $r$ $\leq$ 1$''$ and thus should not contribute
significantly to the counts  within radii $r$ = 1$''$ - 3$''$ shown in this figure.
The strongest excess occurs over the range P.A. = 310$^{\circ}$ - 320$^{\circ}$
(5.8$\sigma$ excess).
}
\end{figure}
\clearpage

%% -------------------------------------------------------
%%%% Fig. 4 RAW and DECON zoom image%%%

\begin{figure}
\figurenum{4}
\epsscale{1.0}
\includegraphics*[width=6.0cm,angle=0]{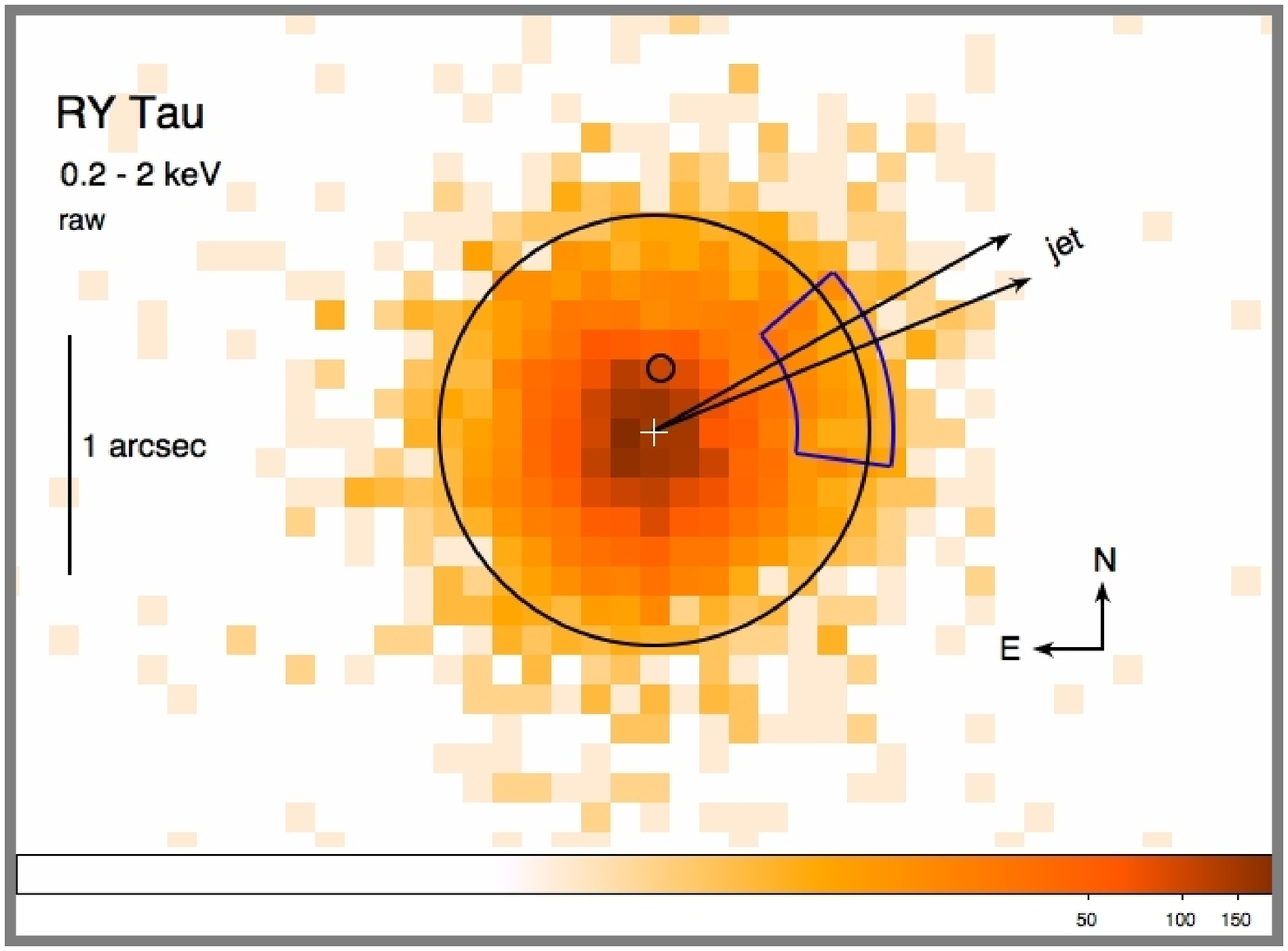} 
\includegraphics*[width=6.0cm,angle=0]{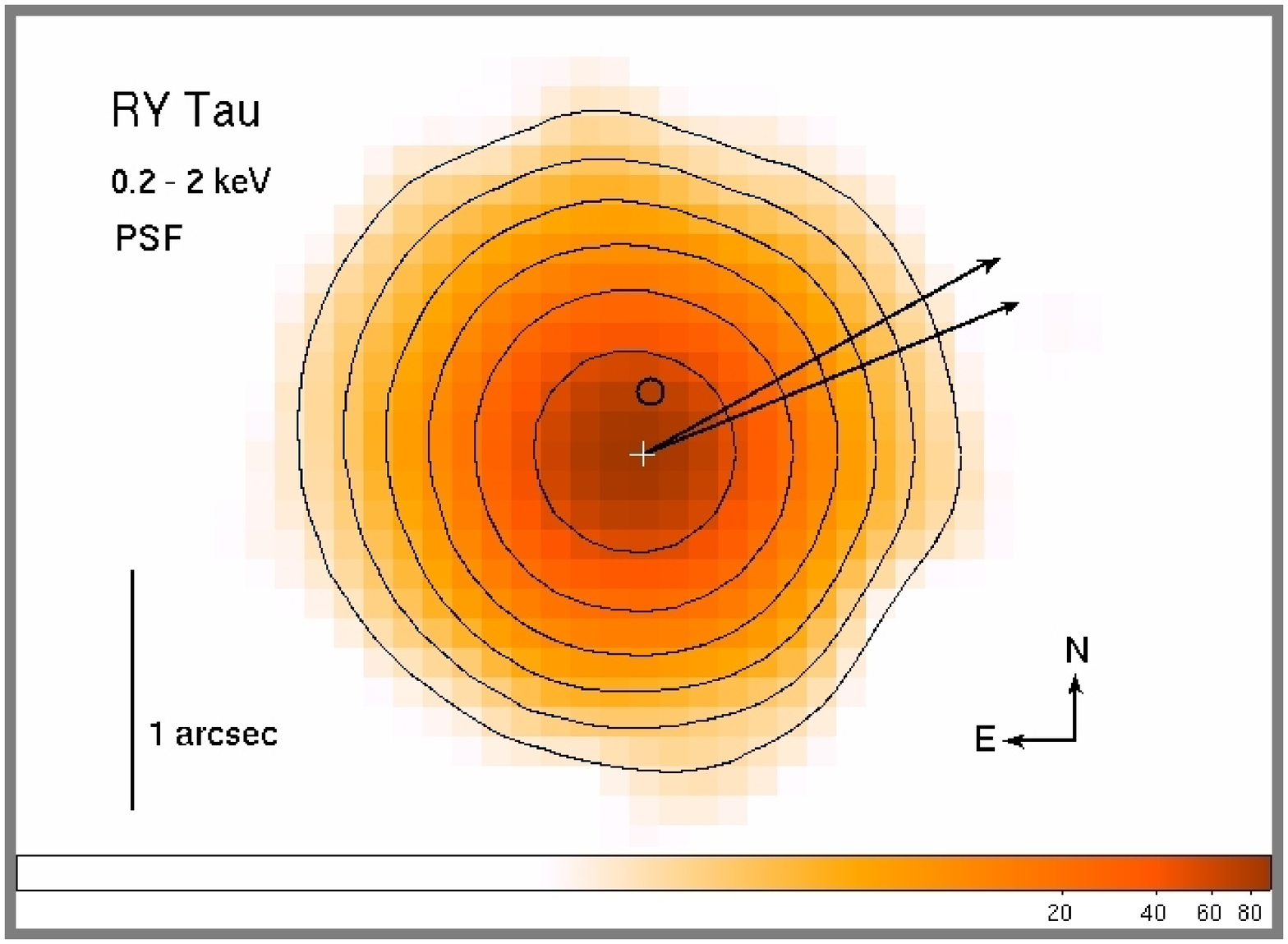} \\
\includegraphics*[width=6.0cm,angle=0]{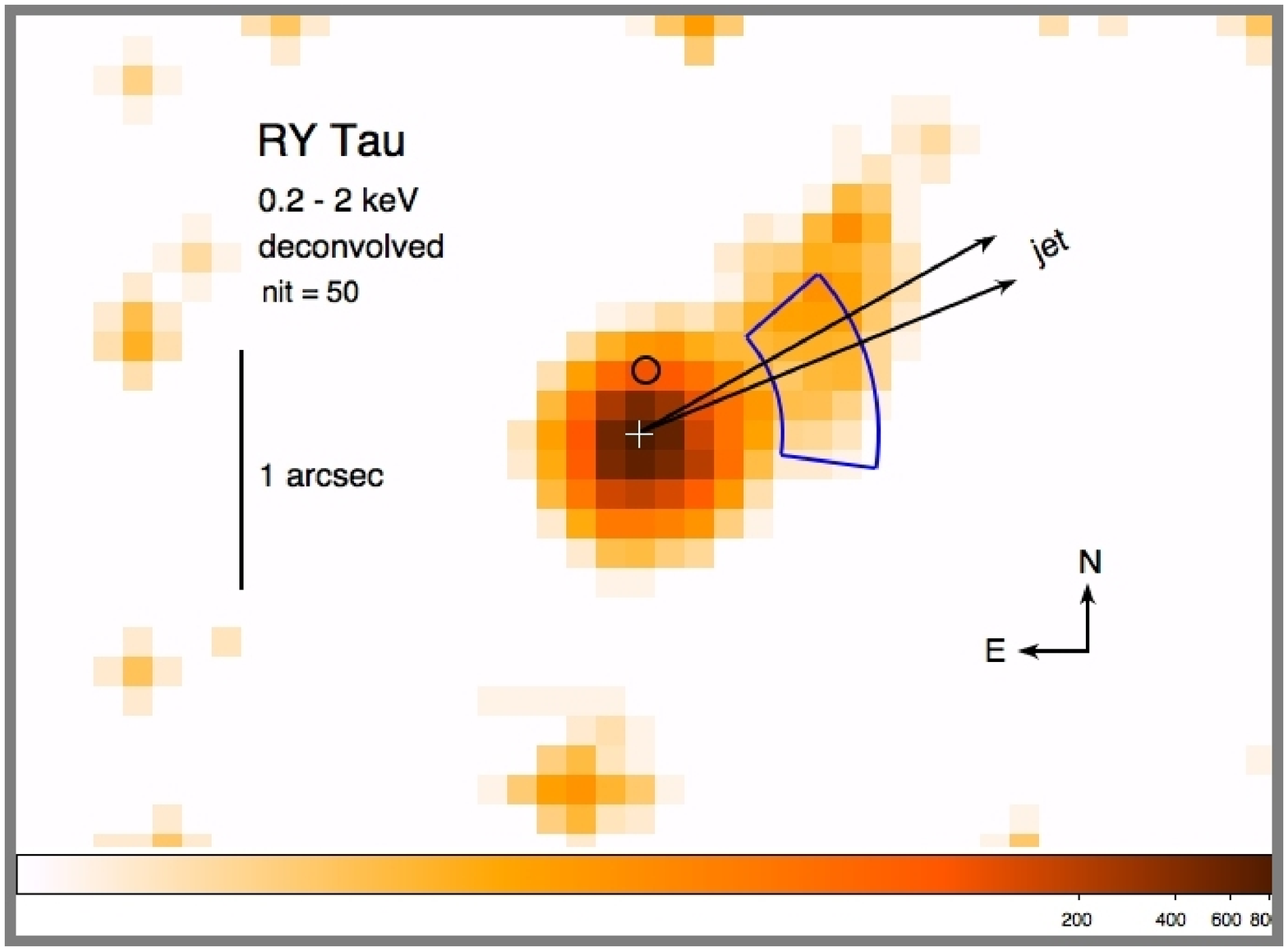} 
\includegraphics*[width=6.0cm,angle=0]{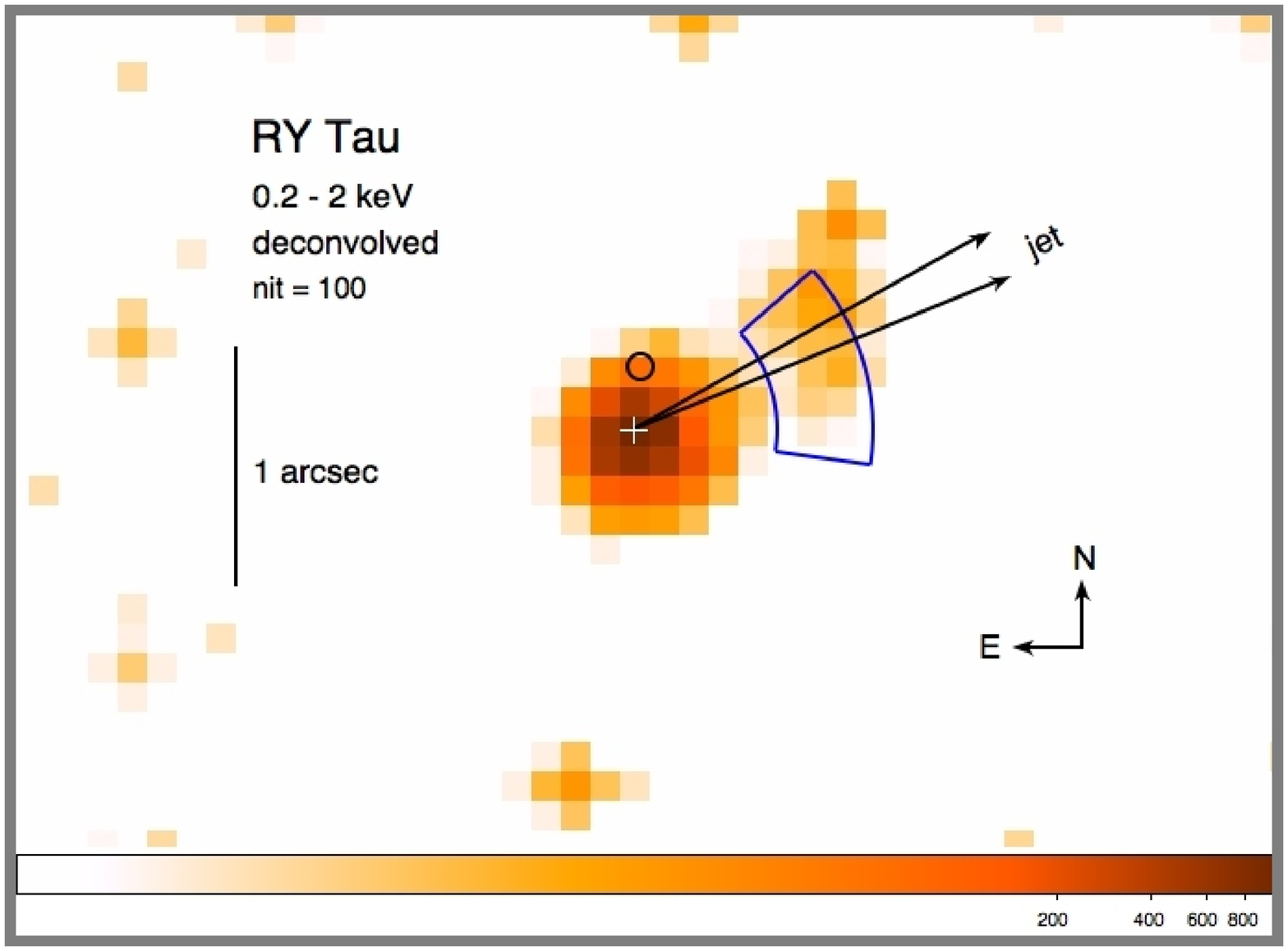} \\
\includegraphics*[width=6.0cm,angle=0]{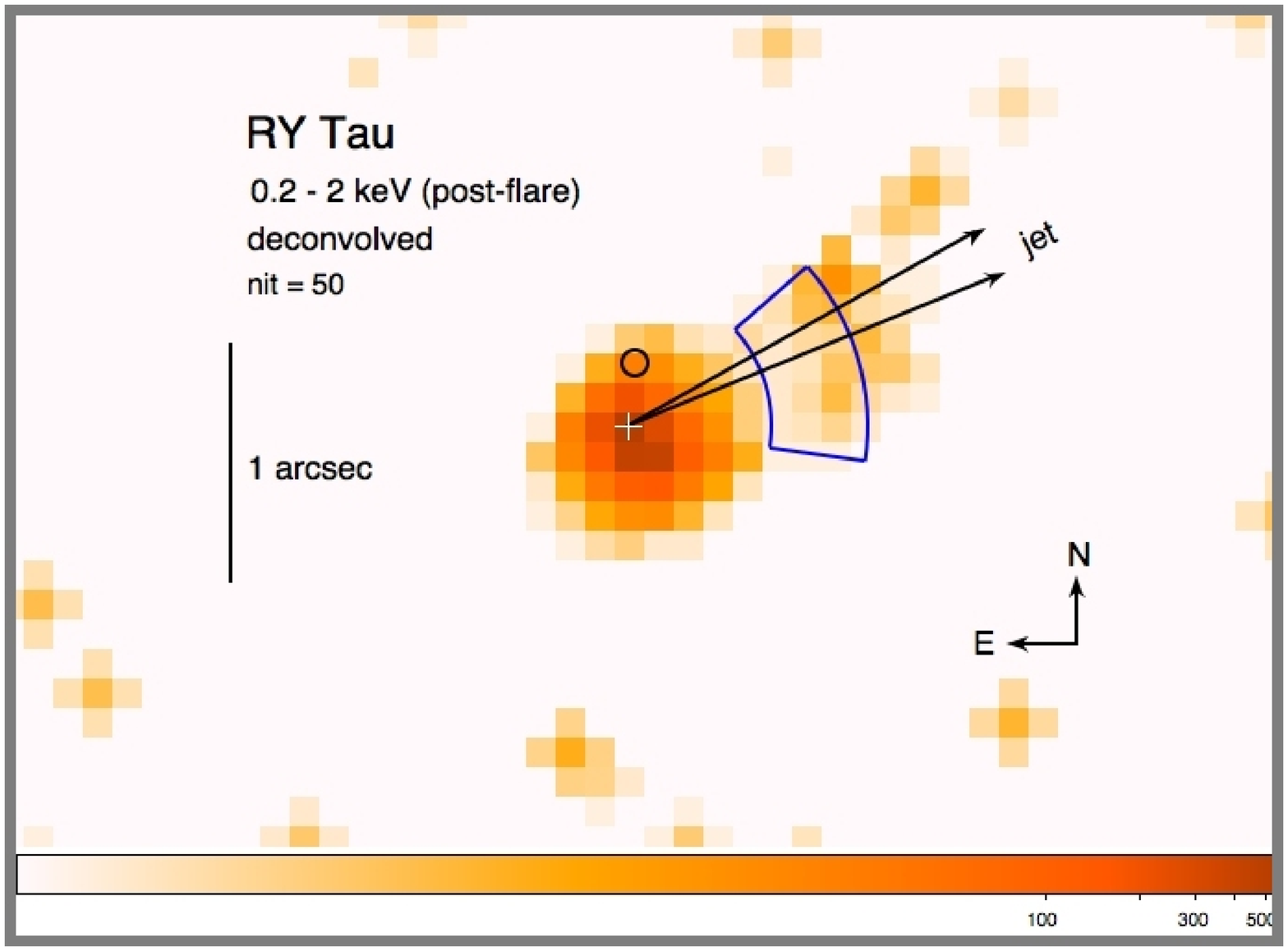}
\includegraphics*[width=6.0cm,angle=0]{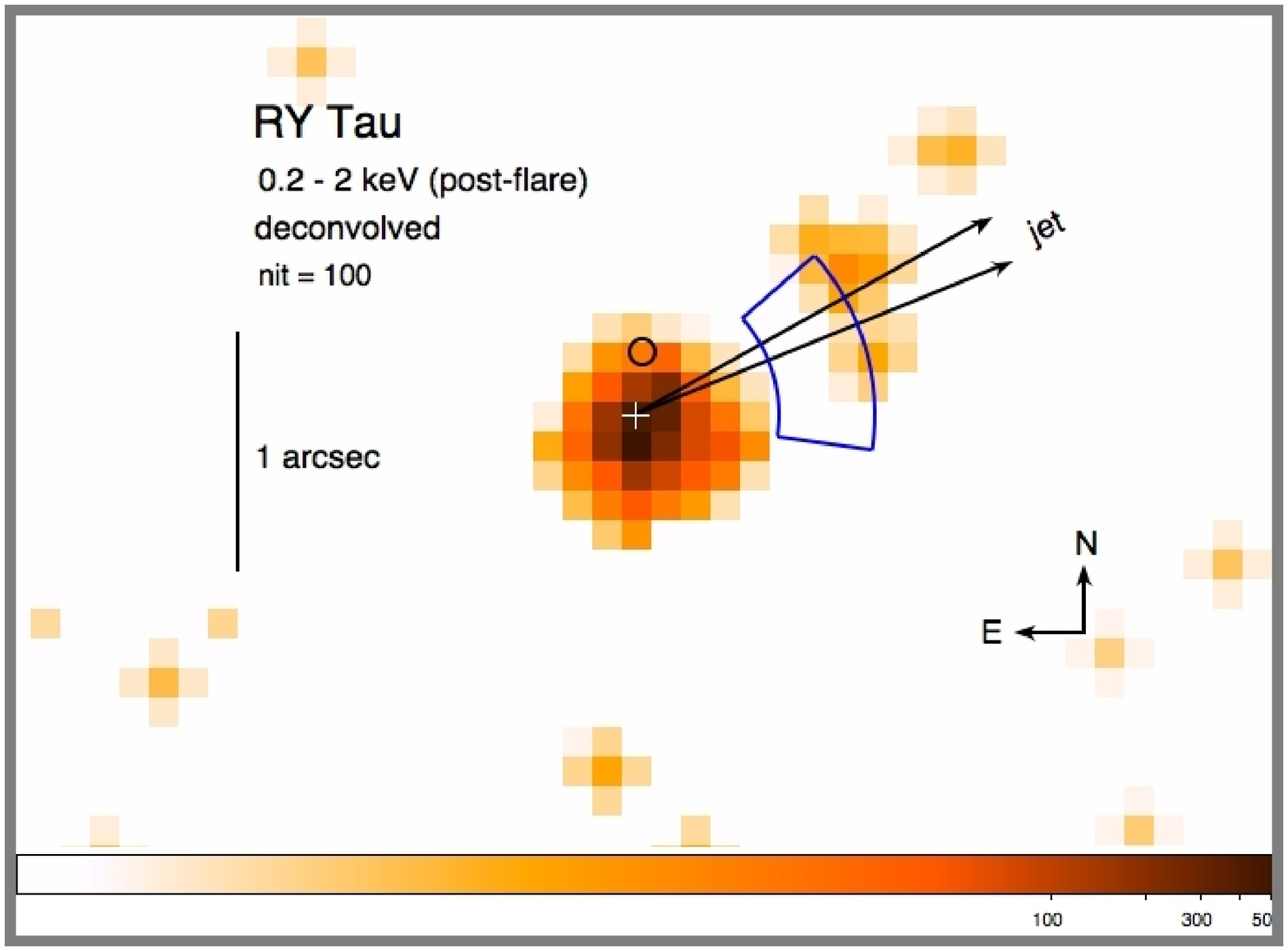} \\
\caption{
\footnotesize{
         Chandra ACIS-S   images of RY Tau in the 
         0.2 - 2 keV band.
         Log intensity scale; pixel size = 0.$''$125; pixel
         randomization removed.
         The cross ($+$) at center marks the  soft-band X-ray
         centroid position (J042157.41$+$282635.22).
         The small circle  marks the {\em HST} GSC vers. 
         2.3.2 position of RY Tau (HST J042157.408$+$282635.48).         
         The vectors of length 1.$''$7 show the range of directions 
         P.A. = 292$^{\circ}$ - 299$^{\circ}$ of the optical jet
         determined from emission knots (SB08).
         The roll angle is 268.6$^{\circ}$ and the focal-plane $+Z$ 
         axis points nearly due west (right). The radially-restricted 
         wedge-shaped region was
         created using the CIAO tool {\em make\_psf\_asymmetry\_region}
         and  encloses 
         the area where the known PSF asymmetry can produce
         artificial structure (P.A. = 286.4$^{\circ}$ $\pm$ 25$^{\circ}$;
         0.$''$6 $\leq$ $r$ $\leq$ 1.$''$0).
         {\em Top Left}:~Raw unsmoothed image using data from the full
           exposure (6013 events in the 0.2 - 2 keV range).  
           The large circle shows the 
           ACIS 90\% encircled power radius  R$_{90}$ = 0.$''$9 (at 
           E $\approx$ 1 keV).~
         {\em Top Right}:~
          PSF image used for deconvolution (full exposure) generated using
          {\em Chart} and {\em MARX}. The image has been Gaussian-smoothed
          using a kernel radius of three pixels. The brightest central pixel
          contains 94 counts.
          The inner contour traces  the PSF half-maximum, which has a radius of
          $\approx$0.$''$4. The outer contour traces the 1-count level.~       
        {\em Middle}:~ Deconvolved  images   
         from CIAO {\em arestore}  using  data from the full
         exposure with 50 iterations (left) and 100 iterations (right).
         The images 
          have been lightly Gaussian-smoothed using a 
         kernel radius of one pixel.~ 
        {\em Bottom}:~ Deconvolved  images, as above, but restricted to
         the post-flare  time interval  ($t$ $>$ 25 ks; 2963 events
         in the 0.2 - 2 keV range).
}
}
\end{figure}
\clearpage
%% ---------------------------------------------------------------

%% -------------------------------------------------------
%%%% Fig. 5 X,Y sky positions %%%

\begin{figure}[h]
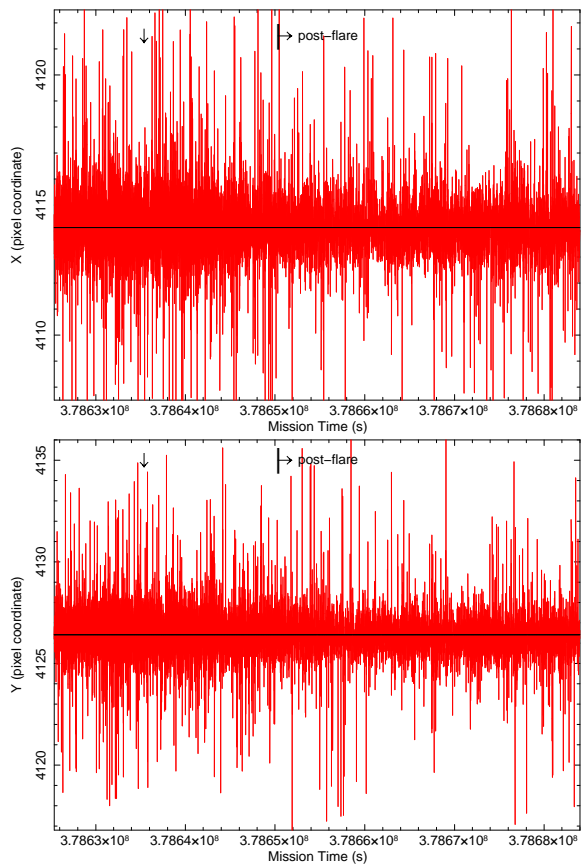

\figurenum{5}
\epsscale{1.0}
\includegraphics*[height=7.7cm,angle=-90]{f5t.eps} \\
\includegraphics*[height=7.7cm,angle=-90]{f5b.eps} 
\caption{Sky-position  of RY Tau as a 
function of mission-time. All events within a radius of 5$''$
from the X-ray peak are shown. No energy filtering has been applied.
The (X,Y) values are sky-pixel coordinates (1 pixel = 0.$''$492)
and correspond to (RA,Dec.)
The downward arrow marks the approximate time of flare peak
(elapsed time $t$ = 10 ks) and the post-flare segment
corresponds to $t$ $>$ 25 ks. The horizontal line is the
mean value. A slightly larger positional scatter 
is seen during the first 25 ks when the flare occurred. 
}
\end{figure}
\clearpage

%% -----------------------------------

%%% Fig. 6 SPECTRUM %%%

\begin{figure}[h]
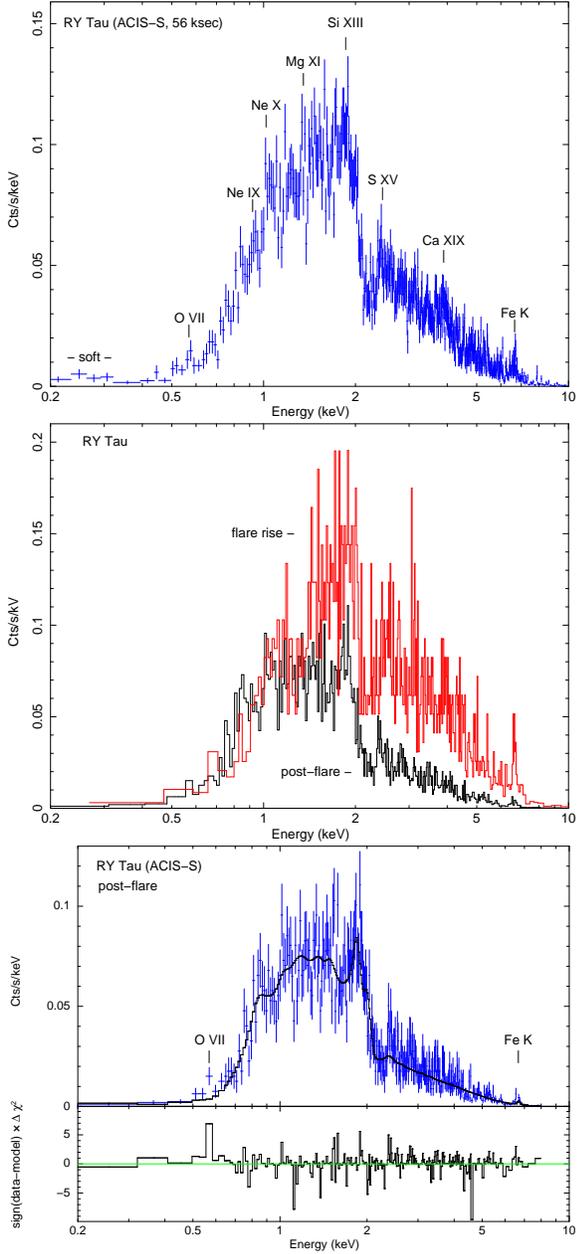

\figurenum{6}
\epsscale{1.0}
\includegraphics*[height=7.7cm,angle=-90]{f6t.eps} \\ 
\includegraphics*[height=7.7cm,angle=-90]{f6m.eps} \\
\includegraphics*[height=7.7cm,angle=-90]{f6b.eps} 
\caption{{\em Top}:~Chandra ACIS-S spectrum of RY Tau  binned to
a minimum of 5 counts per bin. The spectrum was extracted 
from events inside a circular region of radius 3$''$ centered
on RY Tau, and thus includes emission from the star and 
the extended structure to the northwest (Fig. 4). 
The spectrum is  based on the full exposure (12,504 events), 
including events recorded during the flare.   
Faint very-soft emission  in the 0.2 - 0.5 keV range
is present (48 net counts), some of which likely originates 
in the the jet or an accretion shock.
%% *optional spectral plot: preflare vs. postflare
{\em Middle}:~Same as above, except that the spectra were
extracted from specific time intervals corresponding to
the flare rise phase ($t$ = 0 - 7 ks) and late 
post-flare phase ($t$ = 30 ks - 58.6 ks). 
Error bars omitted for clarity.
{\em Bottom}:~Post-flare spectrum overlaid with the best-fit
2T $vapec$ model (Table 2). Note that the model does not
reproduce the faint O VII emission line, suggesting that 
this line may be of non-stellar origin (e.g. arising in
the shocked jet or an accretion shock).
} 
\end{figure}
\clearpage

%%%-----------------------------------

\end{document}